\pdfoutput=1

\documentclass[sigconf]{acmart}

\usepackage{dirtytalk}
\usepackage{siunitx}
\usepackage[inline]{enumitem}
\acmConference[EASE'22]{Evaluation and Assessment in Software Engineering}{June 13--15, 2022}{G\"{o}teborg, Sweden}

\newcommand{\oss}{open-source software}
\newcommand{\perf}{performance}
\newcommand{\spe}{\textsc{SPE}}

\title{The Evolving Landscape of Software Performance Engineering}

\begin{abstract}
Satisfactory software \perf\ is essential for the adoption and the success of a product.
In organizations that follow traditional software development models (e.g., waterfall),
Software Performance Engineering (\spe) involves time-consuming experimental modeling and \perf\ testing outside the actual production environment.
Such existing \spe\ methods, however, are not optimized for environments utilizing
Continuous Integration (\textsc{CI}) and Continuous Delivery (\textsc{CD}) that
result in high frequency and high volume of code changes.
We present a summary of lessons learned and propose improvements to the \spe\ process in the context of \textsc{CI}/\textsc{CD}.
Our findings are based on \spe\ work on products A and B conducted over \num{5} years at an online services company X.
We find that
\begin{enumerate*}[label=(\alph*),before=\unskip{ }, itemjoin={{, }}, itemjoin*={{, and }}]
    \item \spe\ has mainly become a post hoc activity based on data from the production environment
    \item successful application of \spe\ techniques require frequent re-evaluation of priorities
    \item engineers working on \spe\ require a broader skill set than one traditionally possessed by engineers working on \perf.
\end{enumerate*}
\end{abstract}

\begin{document}

\author[1]{Gunnar Kudrjavets}
\orcid{0000-0003-3730-4692}
\affiliation{
   \institution{University of Groningen}
   \city{Groningen}
   \country{Netherlands}}
\email{g.kudrjavets@rug.nl}

\author[2]{Jeff Thomas}
\orcid{0000-0002-8026-9637}
\affiliation{
    \institution{Meta Platforms, Inc.}
    \streetaddress{1 Hacker Way}
    \city{Menlo Park}
    \state{CA}
    \country{USA}
    \postcode{94025}}
\email{jeffdthomas@fb.com}

\author[3]{Nachiappan Nagappan}
\orcid{0000-0003-1358-4124}
\affiliation{
    \institution{Meta Platforms, Inc.}
    \streetaddress{1 Hacker Way}
    \city{Menlo Park}
    \state{CA}
    \country{USA}
    \postcode{94025}}
\email{nnachi@fb.com}



\maketitle

\section{Introduction}

Software systems often not meeting \perf\ requirements is a known problem in the industry~\cite{menasce_2002}.
The accepted definition of \spe\ is \say{a proactive approach that uses quantitative techniques to predict the \perf\ of software early in design to identify viable options and eliminate unsatisfactory ones before implementation begins}~\cite{connie_2015}.
In this paper, we enumerate the lessons learned from our experience while working on tasks related to \spe\ at the company X.
Most of our experience comes from being responsible for the \spe\ of two products (A and B) for \num{5} years.

\emph{Context}.
Company X is a large online services company.
Software developed by X includes
closed-source mobile applications,
\oss\ (such as a memory allocator and a key-value store)
and significant contributions to the Linux kernel.
Product A is an \oss\ database storage engine.
Classic database metrics such as queries per second and transactions per second are the most relevant \perf\ metrics for product A.
Product B is a collection of mobile applications used on various devices running either Android or i{OS}.
Several metrics quantify the performance of mobile applications~\cite{hort_2021}.
The metrics for B include application start time~\cite{verma_optimizing_2015}, memory usage~\cite{lee_2021}, and responsiveness~\cite{yang_2013}.

\emph{Past approaches to \spe}.
In organizations that use a classical waterfall model, the \perf\ work follows a specific model.
Engineers design experiments to measure the \perf\ of a component.
The experiment measures a component's \perf\ either in isolation or in conjunction with the rest of the system.
Based on the results from these experiments,
engineers optimize either design or code until \perf\ metrics meet specific criteria.
A feature that meets the \perf\ criteria is either integrated into the main development branch or released to a subset of users.
The product teams make design and code changes based on the data provided by the \perf\ team.
The rationale for this approach is the assumption that software is modified infrequently once released, and patching is an involved process
(e.g., updating Microsoft Windows).

\section{Industry challenges}

\emph{Disruption by CI/CD}.
Environments that use \textsc{CI}/\textsc{CD} present different constraints for \perf\ engineering than those utilizing the traditional development models.
With the advent of \say{trunk based development}~\cite[p.~339]{winters_2020}, code is no longer developed in hierarchical child branches.
Hundreds to thousands of engineers can commit code to the main branch every day.
\emph{Each commit can potentially cause or fix a \perf\ regression}.
Though we cannot publish a specific number about the ratio of software engineers contributing to the product and \perf\ engineers at X, we can say that the number of product engineers greatly exceeds the number of \perf\ engineers.
Functionality in products is typically enabled or disabled by using \emph{feature flags} (\emph{feature toggles})~\cite{fowler_feature_flags} depending on variables such as deployment goals, target audience, or state of the feature (e.g., GateKeeper~\cite{engineering_at_meta_building_2021}).
Given $N$ feature flags, there is a possibility of $2^N$ combinations of features being active simultaneously.
Testing all these combinations before the release is not practical (or even possible, given that $N$ may be in hundreds).

\emph{Limitations of dogfooding}.
Microsoft popularized the concept of \say{dogfooding} in the software industry~\cite{harrison_2006}.
Dogfooding is the practice of engineers using and testing the software they develop prior to the
release to users.
While this technique has been, in our experience, a valuable tool, we observe that engineers form a biased test sample to detect \perf\ regressions.
Engineers
\begin{enumerate*}[label=(\alph*),before=\unskip{ }, itemjoin={{, }}, itemjoin*={{, and }}]
    \item tend to have better hardware than an average user
    \item have different usage patterns
    \item use the software using outlier configurations settings (e.g., specific feature flags, internal tracing mode, debug builds).
\end{enumerate*}
Dogfooding enables the early identification of obvious \perf\ issues but does not serve as a good predictor for how an average user will behave.

\emph{Increased required skill set}.
Modern software systems are complicated and have many dependencies.
Detecting, debugging, and fixing \perf-related issues in a product requires an
ever-expanding skill set and mastery of different tools~\cite{gregg_systems_2020}.
In our experience, a wide array of knowledge spanning from the internals of operating systems development,
code generation by compilers,
internals of a particular programming language to
product code itself is required.
We observe that debugging and profiling kernel mode and user mode code is necessary to solve complex \perf\ issues.
Assuming that each product team has engineers with that skill set is unrealistic.
Therefore, more requirements are placed on engineers working on \spe.

\section{Method}

Given the relatively small number of engineers working on \spe\ for products A and B, we elect not to perform formal personal opinion surveys.
However, we conducted annual \say{\perf\ summits} (informal focus groups with a mediator) with engineers involved in various \spe-related tasks.
These events, follow-up discussions between engineers, and internal technical white-papers documenting practical \spe\ challenges serve as a basis of this paper.

\section{Findings and Results}

An existing study summarizing \num{9} years of \spe\ experience at {F}idelity states that their \spe\ team was forced to evolve \say{into a rag tag team of system integrators} who were, amongst other things, \say{capable of profiling, instrumenting and changing application code}~\cite{sankarasetty_2007}.
Our experiences with \spe\ are similar to the following findings:

\begin{enumerate}
    \item
     A large part of \spe\ is conducted based on data collected from the production environment (e.g., various counters, logs, snapshots).
    Simulating the production environment in-house is a challenging research problem.
    For example, to correctly evaluate mobile software performance and limit the presence of confounding variables, a Faraday cage may be necessary.

    \item
    A software \perf\ engineer in a \textsc{CI}/\textsc{CD} environment works in a constantly evolving environment given the large volume of code changes due to commits in application code base and related dependencies.

    \item
    Skill set required to be a successful software \perf\ engineer has changed.
    Software \perf\ engineers design experiments, measure their results and present the data to the product teams.
    They are also generalists who can work with a variety of teams and have the ability to debug and profile code at any abstraction level.

    \item
    In our experience, it is valuable to have a small number of dedicated \perf\ engineers working alongside a larger team of product engineers.

    \item
    Classic \perf\ metrics include battery (power consumption), \textsc{CPU}, \textsc{I/O}, and memory~\cite{gregg_systems_2020}.
    Products may demand a different focus and can value one metric over another.
    For example, a long-running process executing on a server in a data center is not concerned as much about the initial process start time as an application executing on a mobile device.

\end{enumerate}

\section{Conclusion}

The role of engineers working in the field of \spe\ for software used daily by millions of people has changed.
The focus has shifted from traditional, primarily theoretical scientific methods (e.g., experimental design, modeling, projections) to a more practical hands-on approach (e.g., debugging, profiling, fixing \perf\ regressions).

Modern software development (e.g., \textsc{CI}/\textsc{CD}) uses \say{release early, release often} type of methodology originating from the early years of the \oss\ movement~\cite{raymond_cathedral_1999}.
The nature of \spe\ has adapted  to accommodate that approach.
Contemporary \spe\ involves working closely with production data and debugging and profiling user and kernel model applications. %

\bibliographystyle{ACM-Reference-Format}
\bibliography{performance}

\end{document}